\documentclass[%
 aip,
 amsmath,amssymb,
 reprint,%
]{revtex4-1}

\usepackage{graphicx}% Include figure files
\usepackage{dcolumn}% Align table columns on decimal point
\usepackage{bm}% bold math
\usepackage{xcolor}
\usepackage{amsmath}
\usepackage{float} % \begin{figure}[H]
\usepackage{physics}
\usepackage[colorinlistoftodos]{todonotes}

\usepackage[utf8]{inputenc}
\usepackage[T1]{fontenc}
\usepackage{mathptmx}
\usepackage{etoolbox}

\makeatletter
\def\@email#1#2{%
 \endgroup
 \patchcmd{\titleblock@produce}
  {\frontmatter@RRAPformat}
  {\frontmatter@RRAPformat{\produce@RRAP{*#1\href{mailto:#2}{#2}}}\frontmatter@RRAPformat}
  {}{}
}%
\makeatother
\begin{document}

\preprint{AIP/123-QED}

\title[]{Deeper-band electron contributions to stopping power of silicon for low-energy ions}
% Force line breaks with \\
%\author{A. Author}
%\author{B. Author}%
% \email{Second.Author@institution.edu.}
%\affiliation{ 
%Authors' institution and/or address%\\This line break forced with \textbackslash\textbackslash
%}%

% \homepage{http://www.Second.institution.edu/~Charlie.Author.}
%\affiliation{%
%Second institution and/or address%\\This line break forced% with \\
%}%

\author{F. Matias}
\email[Authors to whom correspondence should be addressed: ]{phdflaviomatias@gmail.com}
\affiliation{Instituto de Pesquisas Energéticas e Nucleares, Av. Professor Lineu Prestes, São Paulo, 05508-000, Brazil}

\author{P.L. Grande}
\affiliation{Instituto de Física da Universidade Federal do Rio Grande do Sul, Av. Bento Gonçalves, Porto Alegre, 9500, Brazil}

\author{N. E. Koval}
\affiliation{Centro de F\'isica de Materiales, Paseo Manuel de Lardizabal 5, Donostia-San Sebasti\'an, 20018, Spain}

\author{J. M. B. Shorto}
\affiliation{Instituto de Pesquisas Energéticas e Nucleares, Av. Professor Lineu Prestes, São Paulo, 05508-000, Brazil}

\author{T. F. Silva}
\affiliation{Instituto de Física da Universidade de São Paulo, Rua do Matão, trav. R187, São Paulo, 05508-090, Brazil}

\author{N. R. Arista}
\affiliation{División Colisiones Atómicas, Instituto Balseiro, Centro Atómico Bariloche, and Comisión Nacional de Energía Atómica, Bariloche, 8400, Argentina}

\date{\today}% It is always \today, today,

\begin{abstract}
This study provides accurate results for the electronic stopping cross-sections of H, He, N, and Ne in silicon in low to intermediate energy ranges using various non-perturbative theoretical methods, including real-time time-dependent density functional theory, transport cross-section, and induced-density approach. Recent experimental findings [Ntemou \textit{et al.}, Phys. Rev. B {\bf 107}, 155145 (2023)] revealed discrepancies between the estimates of density functional theory and observed values. 
We show that these discrepancies vanish by considering the nonuniform electron density of the silicon deeper bands for ion velocities approaching zero ($v \to 0$).  This indicates that mechanisms such as ``elevator'' and ``promotion,'' which can dynamically excite deeper-band electrons, are active, enabling a localized free electron gas to emulate ion energy loss, as pointed out by [Lim \textit{et al.}, Phys. Rev. Lett. {\bf 116}, 043201 (2016)]. The observation and the description of a velocity-proportionality breakdown in electronic stopping cross-sections at very low velocities are considered to be a signature of the deeper-band electrons' contributions.
\end{abstract}

\maketitle

%\begin{quotation}
%The ``lead paragraph'' is encapsulated with the \LaTeX\ 
%\verb+quotation+ environment and is formatted as a single paragraph before the first section heading. 
%(The \verb+quotation+ environment reverts to its usual meaning after the first sectioning command.) 
%Note that numbered references are allowed in the lead paragraph.
%
%The lead paragraph will only be found in an article being prepared for the journal \textit{Chaos}.
%\end{quotation}

\section{Introduction}

Recent studies on the interaction of various ions with solids in
the low and intermediate energy ranges provide new and relevant insights into 
various aspects of the interaction process, including important
nonlinear and band-structure effects. \cite{pruneda2007,quijada2007,goebl2014,roth2017a,roth2017b,sortica2017,grande16,matias2017,ullah2018}
In particular, recent experimental studies on the stopping power of light
and heavy ions in TiN and Si targets \cite{sortica2019,ntemou2023} have reported significant discrepancies in electronic stopping cross-sections compared to the
standard estimates of density functional theory (DFT). \cite{puska1983,echenique1986}
The reported effect consists of a strong enhancement of the experimental values relative to DFT predictions (assuming uniform electron density conditions).
\cite{sortica2019,ntemou2023}
In a previous study, \cite{matias2019} we provided a quantitative analysis of the experimental results 
reported in Ref.~\cite{sortica2019} for TiN using two alternative theoretical frameworks, namely: 
(i) a self-consistent model for a nonuniform electron gas together with quantum 
scattering and transport cross-section (TCS) calculations, and
(ii) a Penn-type ensemble of electrons also using TCS calculations. 
It is important to note that in both non-perturbative methods, a non-linear description of the
interaction between incident ions and target electrons was applied. 

In the present study, we extend the previous theoretical analysis to the most recent experiments on Si targets, \cite{ntemou2023} showing that these results may also be described in quantitative terms using the previously proposed theoretical approach. \cite{matias2019} In this study, we extend the methodology described in Ref.~\cite{matias2019}, incorporating the induced-density approach and real-time time-dependent density functional theory (TDDFT) to calculate the local electronic stopping power.

Real-time TDDFT is a state-of-the-art methodology for calculating non-perturbative electronic stopping power. However, applying real-time TDDFT to atomistic models can be computationally demanding. Multiple trajectories must be considered to accurately sample all possible electron density regions to obtain a random (or average) stopping power. \cite{gu2020,kononov2023} Moreover, heavy projectile descriptions require the explicit inclusion of core electrons, which play a significant role in the projectile's energy loss. \cite{ullah2018}
Real-time TDDFT applied to uniform electron gas is more computationally efficient; however, it is more suitable for free-electron metals. To describe more complex targets, one has to account for density nonuniformities. 

Combining real-time TDDFT with the Penn model has successfully predicted the electronic stopping power of various targets with high accuracy. \cite{matias:2024} Such an approach incorporates the efficiency and accuracy of real-time TDDFT for a nearly uniform electron gas with the Penn approach describing the electron density in different regions of the target material.

\section{Theoretical procedures}

The theoretical formulation contains some important physical ingredients, as follows:

(1) Nonlinear screening (Friedel sum rule): The interaction of slow ions with solids requires a quantum mechanical analysis even for light ions such as H and He; \cite{puska1983,echenique1986} the nonlinear character of the
electron-ion interactions becomes increasingly relevant for heavier ions. \cite{echenique1986,matias2019}
DFT provides a useful tool for accurately describing the
interaction of target electrons with intruder ions. \cite{puska1983,echenique1986}
An alternative method, first used by Ferrell and Ritchie
\cite{ferrell1977} and followed by Cherubini and Ventura \cite{cherubini1985} and Apagy and Nagy, \cite{apagyi1987}
consists of using analytical models for the screening potential, which contain a screening parameter whose value
is adjusted in a self-consistent way using the so-called Friedel sum rule (FSR). \cite{friedel1952,friedel1954} This rule
was originally derived for the case of static impurities in a free electron gas (FEG); 
an extension of this rule for the case of moving ions was obtained later. \cite{lifschitz1998a} The extended FSR
was applied to light and heavy ions in an extended range of velocities, showing excellent agreement with the experimental values. \cite{lifschitz1998b,arista1999,arista2002}

(2) Local-density approximation (LDA): 
This approximation is long-standing, following a set of pioneering studies by Lindhard and co-workers, \cite{LS1953,Bond1967,Bond1971} who obtained very good agreements with experimental values of stopping powers and energy straggling 
for protons and helium ions in solid targets. Using the LDA method, Rousseau \emph{et al.} \cite{Rousseau1971} successfully explained the
oscillatory dependence of the stopping cross section (SCS) of alpha particles with the target atomic number $Z_2$.
All these studies were performed for the intermediate and high-energy ranges, where linearized models apply. 
However, the case of slow heavy ions in solids is a much more complex problem. Using a combination of FSR with LDA, Calera-Rubio \emph{et al.} \cite{Calera1994} provided a consistent description of the oscillatory dependence of the stopping power on the ion atomic number $Z_1$. 
An important difference between the latter work and the previous ones is that the calculation of Ref. \cite{Calera1994} applies in the non-perturbative range where quantum cross-section analysis must be considered. In contrast, the previous studies \cite{LS1953,Rousseau1971} pertain to the range of linearized models.

(3) Penn algorithm: 
Penn \cite{penn1987} introduced an algorithm to determine the electron inelastic mean free paths (IMFP) employing a model dielectric function. This function can be derived from experimental optical data of the specific material under investigation, utilizing the Lindhard model dielectric function and considering various plasmon energies or electron densities. Furthermore, the same model has been employed to estimate the stopping power of electrons in different materials \cite{SHINOTSUKA201275} and extended to calculate the nonlinear stopping power of ions. \cite{Maarten:2019b} This extension involved using the energy-loss function (ELF) to weigh the contributions from different electron gas components within a statistical ensemble that characterizes the material of interest.

In this work, we combine the LDA and Penn algorithms for describing non-local electron density with three non-perturbative methods for calculating the electronic stopping power, namely real-time TDDFT, TCS, and IDA, as described below. 

%More recently, time-dependent DFT (TDDFT) methods were developed, with applications to light ions \cite{pruneda2007,ullah2018}. 

\subsection{Real-time TDDFT-, TCS- and IDA-LDA approaches}

The first approach considered here is based on the method proposed in Ref. \cite{matias2019}, which consists of using the LDA method to describe the nonuniform electron density in the Si target, together with nonlinear calculations of the energy loss of ions in a uniform (locally) electron gas. The target
is described by its local density $n(r)$ or equivalently local $r_s(r)$ values, related by  $1/n(r)=(4\pi/3)r_s^3$. 

For calculations using the LDA method, prior knowledge of the electronic density of the Si target is necessary. In this report, the electron density of Si was extracted from the Ziegler tables. \cite{ziegler:1985} The atomic density of Si is $\rho = 2.33$ g/cm$^3$. For the sake of reliability,
we analyzed the normalization of the electron density to the
number of electrons in the Si atom, calculated as

\begin{equation}
    N_e = 4\pi \int_0^{r_{\text{cell}}}n(r)r^2dr.
\label{eq:elec-numb}
\end{equation}

\begin{figure}[h]
\begin{center}
\includegraphics[angle=0,width=8.65cm,clip]{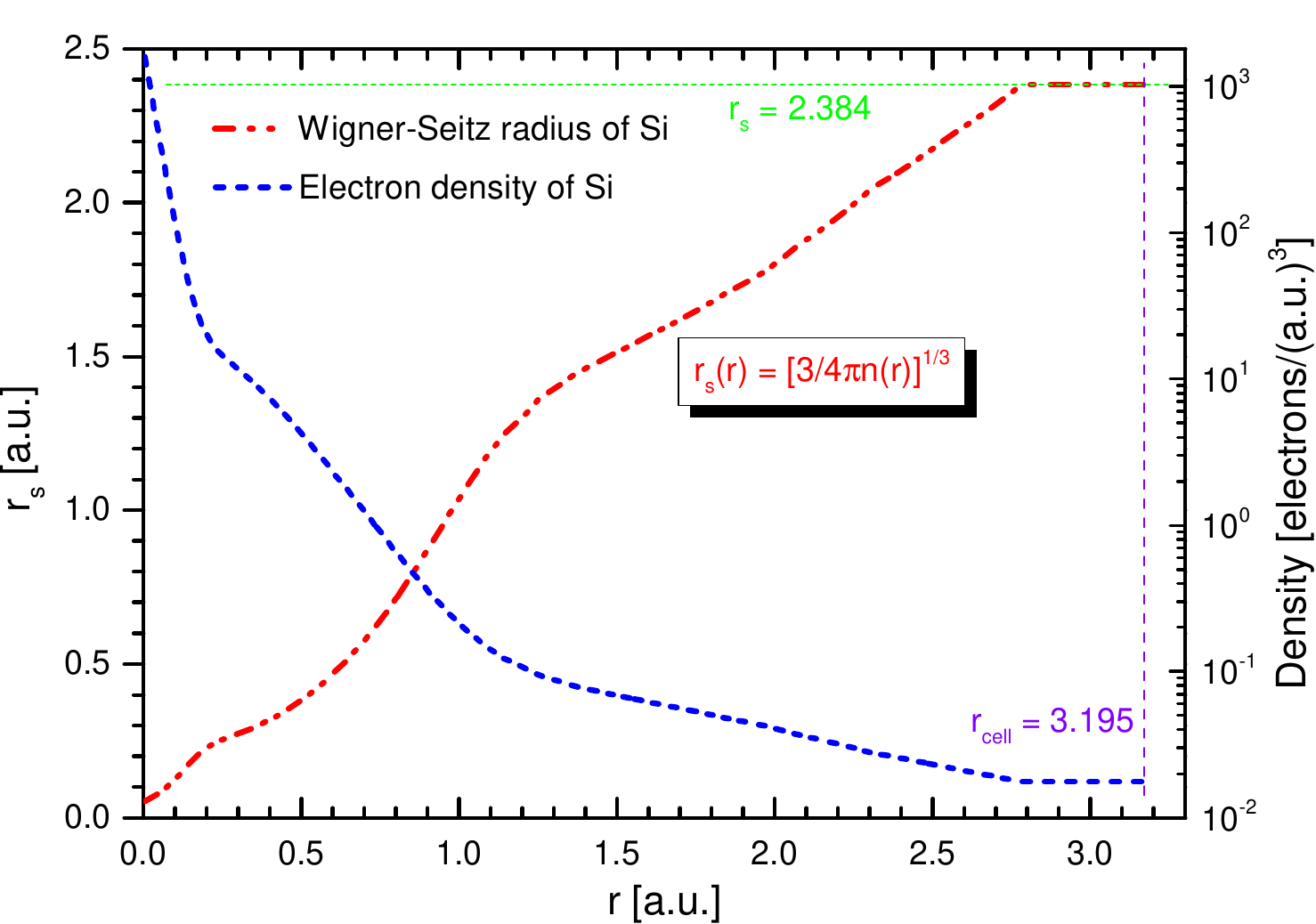}
\caption{Radial electron density (blue dashed line) within the WS sphere encompassing Si atom. The radial distance $r_{\text{cell}}$ = 3.195 a.u. represents the atomic cell radius of Si. The several WS radii are shown in a red dotted line. In cell bord, $r_s = 2.384$ a.u.}
\label{fig:nr-rs_r}
\end{center}
\end{figure}

Equation (\ref{eq:elec-numb}) was applied assuming an atomic Wigner-Seitz (WS) sphere around the atom of Si. The solution of this equation was obtained integrating until the atomic cell radius $r_{\text{cell}}$ = 3.195 a.u. (vertical dashed violet line), resulting in $14.017$ electrons. Figure~\ref{fig:nr-rs_r} shows the electron density concerning the radial distance $r$ from Si nuclei (blue dashed line) and WS radii (red dotted line), $r_s$(r).  

In this approach, the final step is to integrate the stopping powers over the WS cells corresponding to the Si atom using the following expression:

\begin{equation}
    \left[\frac{dE}{dz}(v)\right]_{\text{X-LDA}} = 4\pi N_a \int_0^{\infty}r^2 dr \left[\frac{dE}{dz}(v)\right]_{\text{X}}(v,r_s(r)),
    \label{eq:lda}
\end{equation}

\noindent
where $N_a= 1/V_a$ is the number of Si atom per unit volume: $V_a=(4\pi/3) r_{\text{cell}}^3$. In Equation (\ref{eq:lda}), X $\equiv$ TDDFT, TCS, or IDA. The values of $r_s(r)$ are shown in Fig.~\ref{fig:nr-rs_r}.

\begin{figure}[H]
\begin{center}
\includegraphics[angle=0,width=8.65cm,clip]{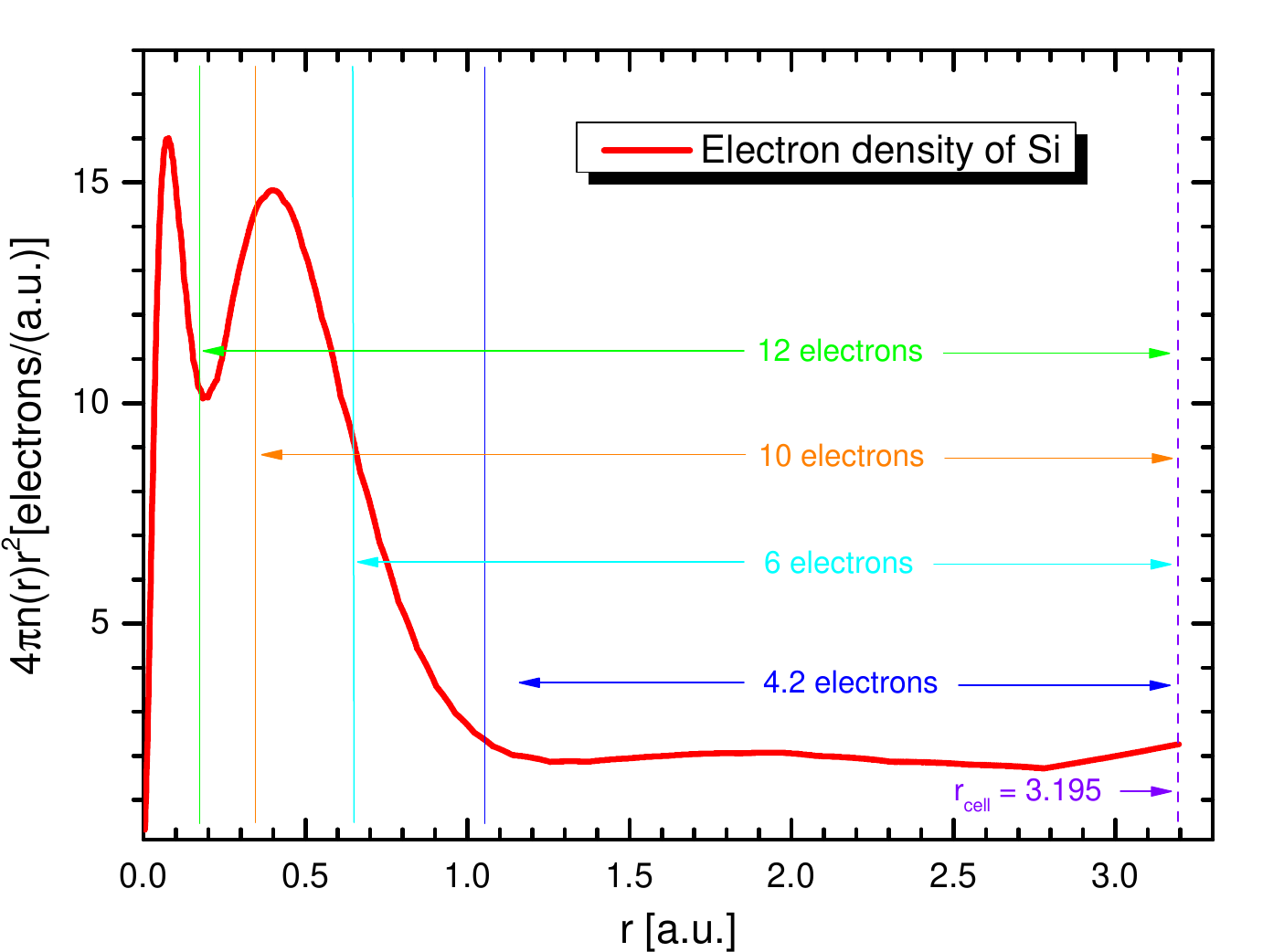}
\caption{Electron density multiplied by the solid angle ($4\pi r^2$) as a function of the atomic
cell radii of Si. The spacing between each solid vertical line and the dashed vertical line (atomic cell
radius of Si, $r_{\rm cell}$) indicates the number of electrons that effectively contribute to the nonuniform FEG in electron-projectiles interaction, namely: $6$ for H$^0$, $8$ for He$^0$, and $10$ for N$^0$ and Ne$^0$. These ranges correspond to converged values of the stopping power obtained using Eq.~(\ref{eq:lda}) in $v=0.1$ a.u. For comparative purposes, we show the radial interval corresponding to $4.2$ electrons (nonuniform FEG), in analogy to the uniform FEG ($4.2$ electrons) with the density parameter $r_s = 1.97$ a.u.}
\label{fig:electrons-number}
\end{center}
\end{figure}

Electron density multiplied by the solid angle ($4\pi r^2$) as a function of the atomic cell radii of Si is shown in Fig.~\ref{fig:electrons-number} (solid line). The vertical solid lines represent the integration ranges for the calculations converged of the Eq.~(\ref{eq:lda}). The knowledge of these intervals was obtained from the convergence of Eq.~(\ref{eq:lda}) in $v=0.1$ a.u. As discussed later, these ranges correspond to the deeper- and valence-band electrons that effectively contribute to the electron-projectiles interaction, namely: $6$ electrons for H$^0$ (0.65 to 3.195 a.u.), $10$ for He$^0$ (0.35 to 3.195 a.u.), and $12$ electrons for N$^0$ and Ne$^0$ (0.17 to 3.195 a.u.). For comparative purposes, in Fig.~\ref{fig:electrons-number}, we also show the radial interval corresponding to $4.2$ electrons, in analogy to the uniform FEG ($4.2$ electrons) with the density parameter $r_s = 1.97$ a.u., a value calculated from the experimental plasmon frequency, $\omega_p$.~\cite{brandt:1981}     

\subsection{Real-time TDDFT-, TCS- and IDA-Penn approaches}

Recently, a new non-perturbative method has been introduced to characterize the electronic stopping power of light and heavy ions in materials. \cite{Maarten:2019b} The Penn approach has been successfully utilized in the TCS for low-energy protons with velocities lower than the Fermi velocity ($v < v_F$). \cite{matias2019} This approach has been successfully applied to predict the accurate electronic stopping power of protons in polymers using real-time TDDFT over a wide range of proton energies ($0.25$ to $10000$ keV).\cite{matias:2024} This method considers the combination of electron gas responses characterized by nonuniform densities. It is similar to the approach outlined in the Penn method. \cite{penn1987} To analyze each free electron density, the material's ELF at the optical limit is used:

\begin{eqnarray}
       g(\omega_p) =\frac{2}{\pi \omega_p}\text{ELF}(\omega_p).
    \label{eq:elf} 
\end{eqnarray}

\begin{figure}[H]
\begin{center}
\includegraphics[angle=0,width=9cm,clip]{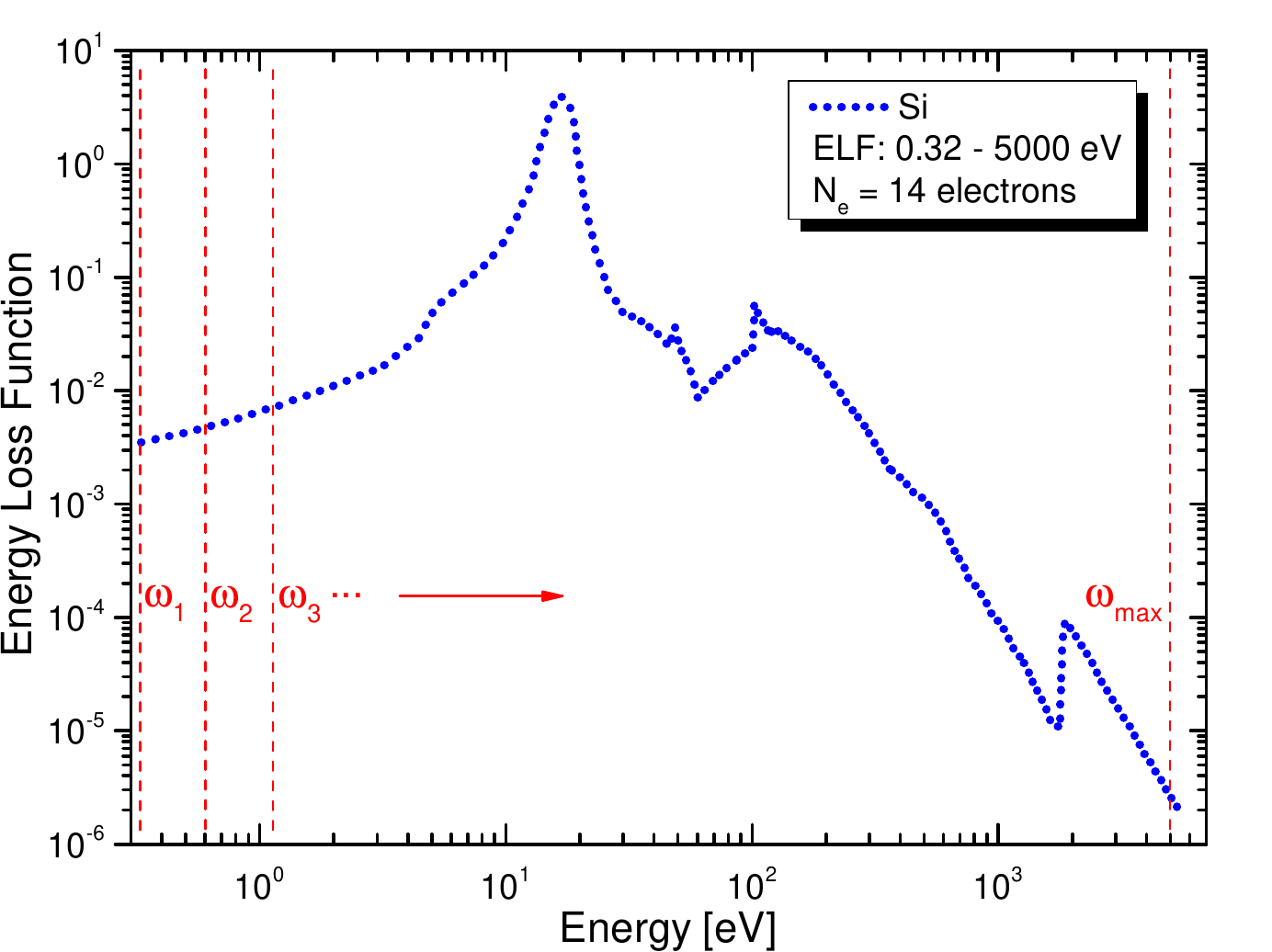}
\caption{Optical-ELF data for Si obtained from.~\cite{tanuma-penn1993} The intervals from $\omega_1$ to $\omega_{\rm max}$ yield 14 electrons from applying the sum rule. \cite{Maarten:2019b}}
\label{fig:elf}
\end{center}
\end{figure}

\noindent
Fig.~\ref{fig:elf} shows the optical-ELF data of Si, as presented in \cite{tanuma-penn1993}. %To achieve convergence of Equation (\ref{eq:penn}), it is necessary to use $d\omega = 0.001$. 
The sum rule \cite{Maarten:2019b} gives 14 electrons.

The stopping power depends on the plasmon frequency $\omega_p$ determined by the contributions of individual electron gases obtained from $r_s$: $\omega_p=\sqrt{3}r_s^{-3/2}$. Hence, the method for calculating the stopping power is as follows:

\begin{equation}
    \left[\frac{dE}{dz}(v)\right]_{\text{X-Penn}} = \int_0^{\infty}d\omega_p g(\omega_p)\left[\frac{dE}{dz}(v)\right]_{\text{X}}(v,\omega_p).
    \label{eq:penn}
\end{equation}

\noindent
where X $\equiv$ TDDFT, TCS, or IDA; the electronic stopping $\displaystyle \left[\frac{dE}{dz}(v)\right]_{\text{X}}$ is given by Eqs.~(\ref{stp_tddft}) and (\ref{eq:stopping-geral}),
%and (\ref{stp_tcs}), and (\ref{eq:stopping-geral}) and (\ref{stp_ida})
respectively (see sections \ref{sec:tddft} and \ref{sec:ida_tcs}). Due to their unique characteristics, we have named these approaches real-time TDDFT-Penn, TCS-Penn, and IDA-Penn.

\subsection{Real-time TDDFT method}\label{sec:tddft}

Real-time TDDFT is a highly accurate \emph{ab initio} tool for describing the electronic stopping power in spherical jelliums. The jellium model assumes a positive background (representing the ion cores) that provides a charge balancing for the electron gas. Compared to fully atomistic models, the advantage of jellium representation is the computational efficiency. The real-time TDDFT in a FEG has shown accurate results for near-free-electron systems. 

The approach adopted in this work reflects the methodology used in Refs., \cite{borisov:2001,borisov:2006,chulkov:2006,quijada2007,quijada10,koval12,koval2013,matias2017} and will be briefly explained in this section. In this approach, the time evolution of the electron density incorporates, in a non-perturbative manner, the complete dynamic interaction between an external field and the medium. This computational framework has been used to analyze various issues in condensed matter systems, such as dynamic charge screening in metallic media, \cite{borisov2004} energy loss of atomic particles in matter, \cite{quijada2007,matias2017} as well as many-body effects associated with hole screening in photoemission. \cite{koval12}

A static density functional theory (DFT) calculation is performed to obtain the system's ground state. The time evolution of the complete electron density, $n({\bf r},t)$, in response to an external field (in this case, a proton), is conducted within the framework of real-time TDDFT in the Kohn-Sham scheme.%~\cite{koval2013,borisov:2006,borisov:2001,chulkov:2006}. 
%The numerical procedure is employed in Refs. \cite{borisov2004,pruneda2007,quijada2007,quijada10}, where additional details can be found.

The energy loss is calculated by integrating the time-dependent induced force $F$ over the proton:
\begin{equation}
E_{\rm{loss}}(v)=-v\int_{-\infty}^{+\infty} F (t)dt,
\end{equation}

\noindent
where $v$ is the (constant) velocity at which the proton traverses the jellium. Once the induced force on the proton is calculated, the average (or effective) stopping power is computed as the energy loss per unit path length, i.e.,
\begin{equation}
\left[\frac{dE}{dz}(v)\right]_{\text{TDDFT}}=\frac{E_{\rm{loss}}(v)}{2R_{cl}}.
\label{stp_tddft}
\end{equation}

Real-time TDDFT is applied only to the electronic stopping power calculations for protons in matter because the current code considers only bare ions as projectiles. Extending this method to heavier ions, including charge states in the electron-ions process, is a complex task requiring substantial effort. Therefore, self-consistent methods based on the FSR and analytical screening potentials currently represent the most convenient approach for studying nonlinear effects in the electronic screening charge and energy losses of heavier ions in matter.

\subsection{IDA and TCS methods}\label{sec:ida_tcs}

Within the framework of stationary states, the electronic stopping power is also calculated using two theoretical models: the TCS \cite{lifschitz1998b} and the induced-density approach (IDA). \cite{grande16,matias2017} 

TCS method was first introduced by Finnemann and Lindhard \cite{finnemann1968} and was applied by Briggs and Pathak to explain the oscillatory dependence of the stopping power on the atomic number $Z_1$ for channeled ions with low velocities. \cite{briggs1973} Significant contributions were then made by Echenique, Puska, Nieminen, Ashley, and Ritchie following the
development of DFT methods. \cite{puska1983,echenique1986}
As indicated before, a simplified nonlinear approach was proposed, which uses parametric screening potentials in conjunction with the FSR. The extension of the FSR to finite ion velocities \cite{lifschitz1998b} opened the way to self-consistent calculations for light and heavy ions in a wide range of velocities.

The TCS \cite{lifschitz1998b} and IDA \cite{grande16,matias2017} calculations use a model potential with parameter $\alpha$ as detailed in Ref.. \cite{matias2017} Through numerical integration of the radial Schrödinger equation, the scattering phase shifts $\delta_\ell(v)$ were determined for a large number of $\ell$ values, depending on both the density parameter $r_s$ and the potential parameter $\alpha$. \cite{matias2017} The value of $\alpha$ (for each $r_s$) was iteratively determined to satisfy the FSR, 

\begin{equation}
    Z=\frac{2}{\pi}\sum_{\ell=0}^{\ell_{\text{max}}}(2\ell + 1)\delta_\ell (v),
\end{equation}

\noindent
requiring multiple iterations for a self-consistent solution. We derived the ultimate phase shift values for each combination of $Z$, $r_s$, and $v$ through an iterative process. The range of $r_s$ values considered was from $0.1$ to $2.384$ a.u., which covered the relevant range of the study (see Fig.~\ref{fig:nr-rs_r}). Once we obtained the $\delta_\ell$ values, we used them to compute the TCS, using the following equation:

\begin{equation}
    \begin{split}
        \left[\frac{dE}{dz}(v)\right]_{\text{X}} &=n_0m_e\left\langle \frac{|\Vec{v}_e-\Vec{v}|}{v}\Vec{v}\cdot (\Vec{v}-\Vec{v}_e)\sigma_{\text{tr}}^{\text{X}} (|\Vec{v}_e-\Vec{v}|) \right\rangle_{\Vec{v}_e},
    \end{split}
    \label{eq:stopping-geral}
\end{equation}

\noindent
where $m_e$ is the electron mass, $\langle \cdot\cdot\cdot \rangle_{\Vec{v}_e}$ stands for the average over the electron velocities $\Vec{v}_e$, $\Vec{v}$ is the ion velocity, $n_0$ is the undisturbed electron density, $\sigma_{\text{tr}}^{\text{X}}$ is the transport cross-section, and $\text{X} \equiv$ IDA or TCS. For $\text{X} \equiv$ TCS, the $\sigma_{\text{tr}}^{\text{TCS}}$ can be expressed by phase shifts $\delta_\ell$ at the relative speed $v^\prime$, according to \cite{psigmund2006}

\begin{equation}
    \sigma_{\text{tr}}^{\text{TCS}}(v^\prime) = \frac{4\pi}{v^{\prime 2}}\sum_{\ell=0}^\infty (\ell+1)\sin^2{[\delta_\ell(v^\prime)-\delta_{\ell+1}(v^\prime)]}.
    \label{stp_tcs}
\end{equation}

For $\text{X} \equiv$ IDA, the electronic stopping power of ions is estimated by Eq.~(\ref{eq:stopping-geral}) with the effective transport cross-section, according to \cite{grande16}

\begin{equation}
    \sigma_{\text{tr}}^{\text{IDA}}(v^\prime) = \frac{2\pi Z}{v^{\prime 3}}\sum_{\ell=0}^\infty\sin{(2 [\delta_\ell(v^\prime)-\delta_{\ell+1}(v^\prime)])}.
    \label{stp_ida}
\end{equation}

\noindent
We highlight that the IDA and TCS methods yield the same results for neutral projectiles. For more details on the IDA model, please refer to the references \cite{grande16,matias2017}.

\section{Results}

\begin{figure}[h!]
\begin{center}
\includegraphics[angle=0,width=8.5cm,clip]{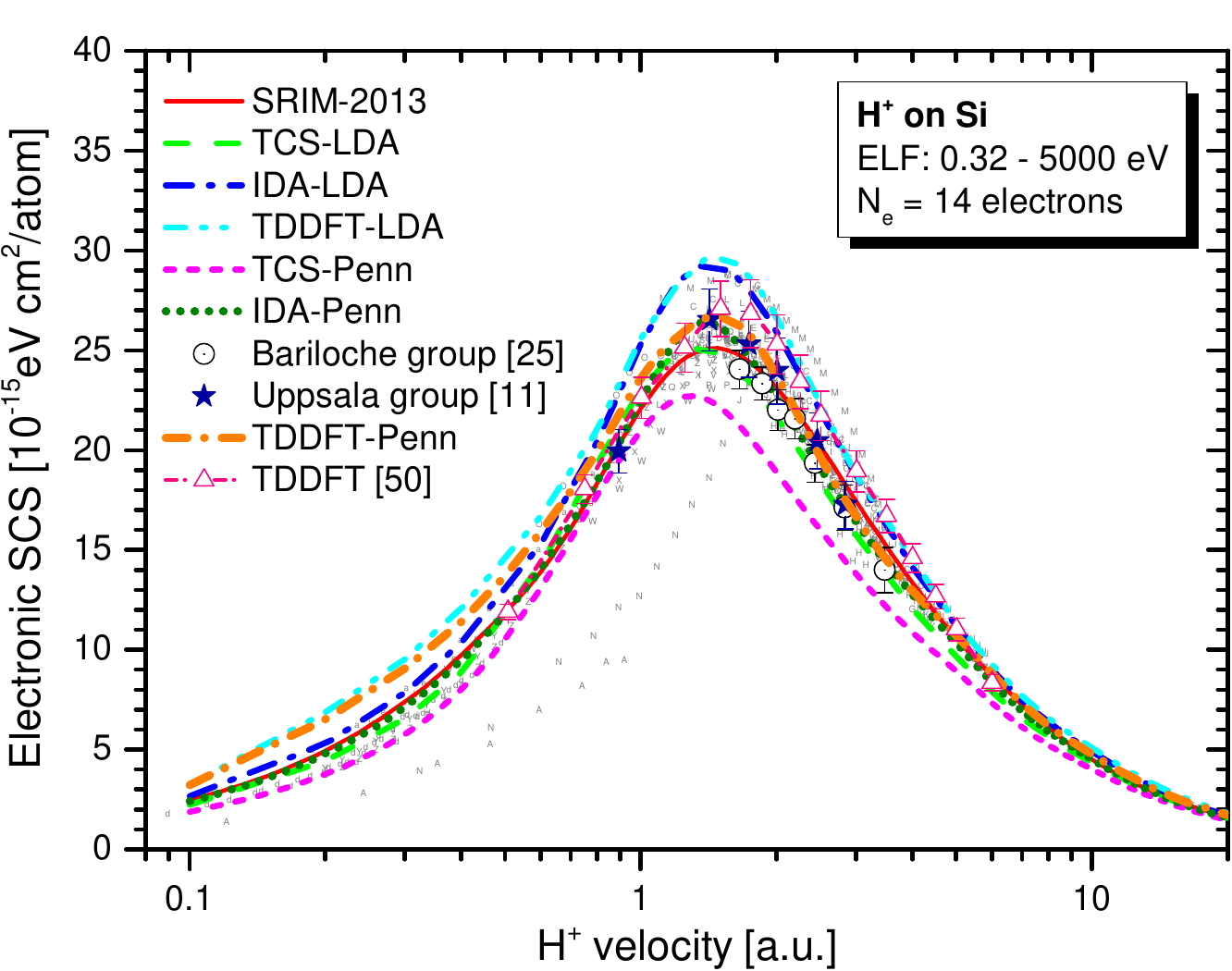}
\caption{Electronic SCS for H$^+$ ions on Si as a velocity function. TCS-LDA (green dashed line), IDA-LDA (blue dash-dotted line), real-time TDDFT-LDA (cyan dash-dot-dotted line), TCS-Penn (magenta short-dashed line), IDA-Penn (olive short-dotted line), and real-time TDDFT-Penn (orange short-dash-dotted line) results are shown. Our results are compared to the semi-empirical SRIM-2013 model (red solid line), \cite{srim2010} and experimental data available at the IAEA database (uppercase letters).~\cite{iaea} Recent data from the Uppsala group (royal star symbols) \cite{ntemou2023} and the experimental data from the Bariloche group (black open circle symbols)~\cite{arista1999} are shown as well. Real-time TDDFT results by Yost \emph{et el.}, \cite{kanai:2017} scaled to account for the core correction, are also presented (light-red empty triangles; line is shown to guide the eye).}
\label{fig:H_Si_10000keV_Penn}
\end{center}
\end{figure}

\begin{figure}[h]
\begin{center}
\includegraphics[angle=0,width=9cm,clip]{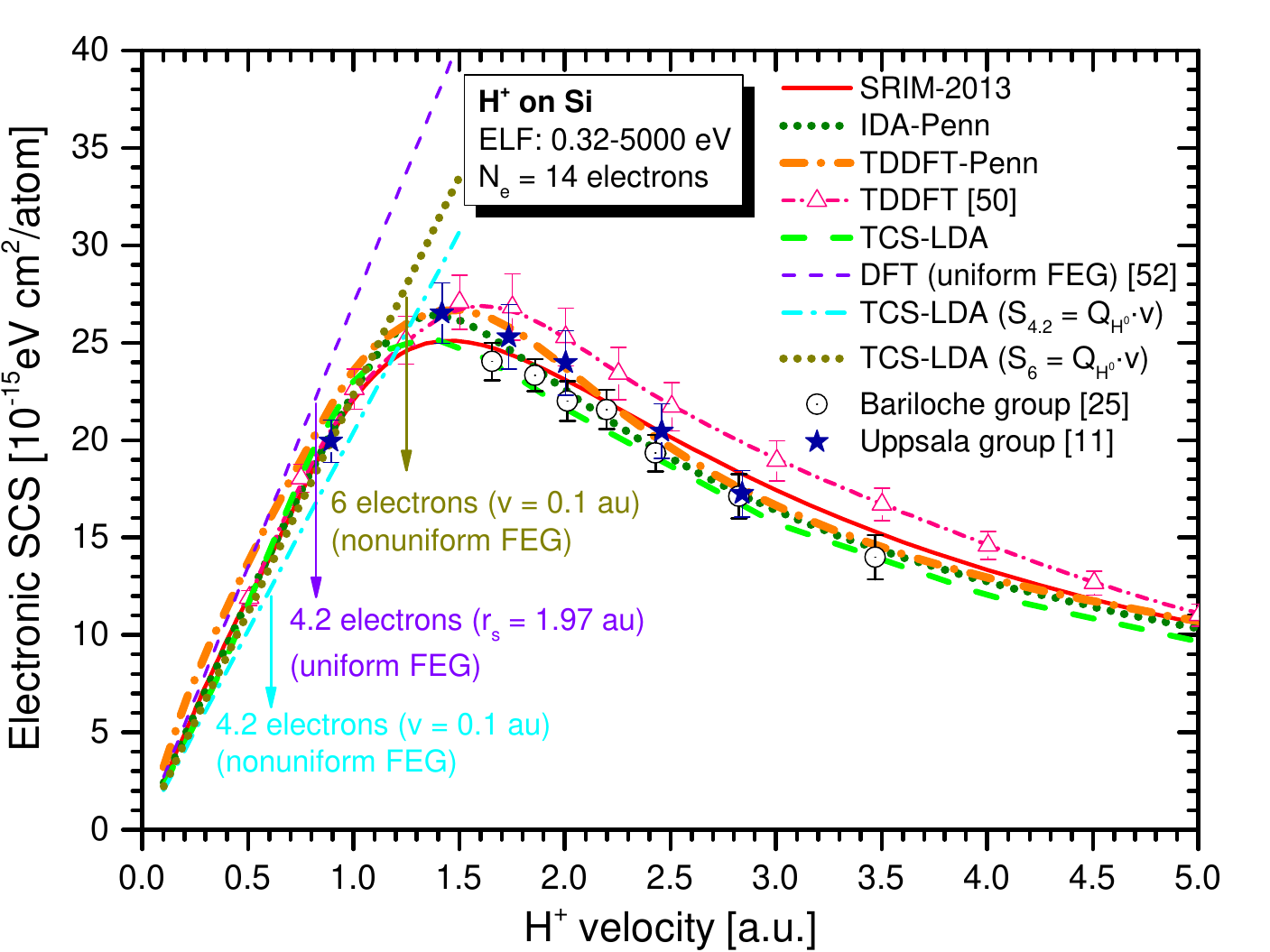}
\caption{As in Fig.~\ref{fig:H_Si_10000keV_Penn}, electronic SCS for H$^+$ ions on Si as a function of velocity is presented. IDA-Penn (olive short-dot line), real-time TDDFT-Penn (orange short-dash-dot line), and TCS-LDA (green-dash line) results are shown. For the comparison proposal, we have plotted the original uniform nonlinear DFT estimates for the linear SCS~\cite{ECHENIQUE1990229}. SCS proportional to the H$^0$ velocity ($S=Q\cdot v$), calculated from the TCS-LDA approach, is shown for case two: cyan dashed dot line is the result, which 4.2 electrons per Si molecule contribute to the nonuniform FEG (see Fig.~\ref{fig:electrons-number}); dark yellow short-dot line is the result with full ELF, in which 6 electrons effectively contribute to the nonuniform FEG in electron-H$^0$ interaction. Real-time TDDFT results by Yost \emph{et el.} \cite{kanai:2017}, scaled to account for the core correction, are also presented (light-red empty triangles; line is shown to guide the eye).}
\label{fig:H_Si_625keV_LDA}
\end{center}
\end{figure}

\begin{figure}[H]
\begin{center}
\includegraphics[angle=0,width=9cm,clip]{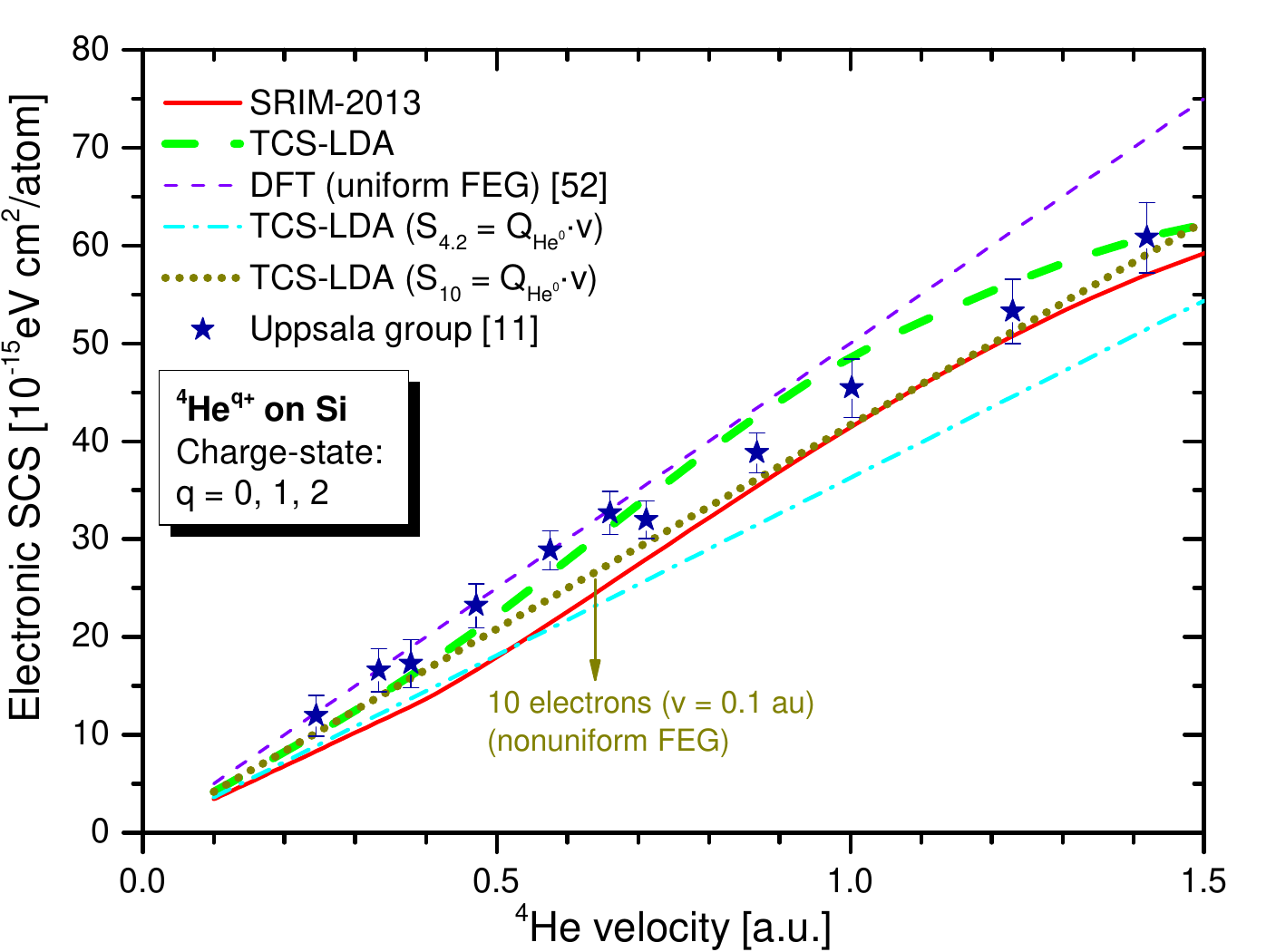}
\caption{Electronic SCS averaged over the charge states for He$^{\text{q}+}$ ions on Si as a velocity function. TCS-LDA (green-dash line) results are shown. The original uniform nonlinear DFT estimates,~\cite{ECHENIQUE1990229} proportional to the He$^0$ velocity. As in Fig.~\ref{fig:H_Si_625keV_LDA}, TCS-LDA for the linear SCS is presented: the dark yellow short-dot line is the result with full ELF, in which 10 electrons effectively contribute to the nonuniform FEG in electron-He$^0$ interaction.}
\label{fig:He_Si_56keV_LDA}
\end{center}
\end{figure}

\begin{figure}[h]
\begin{center}
\includegraphics[angle=0,width=9cm,clip]{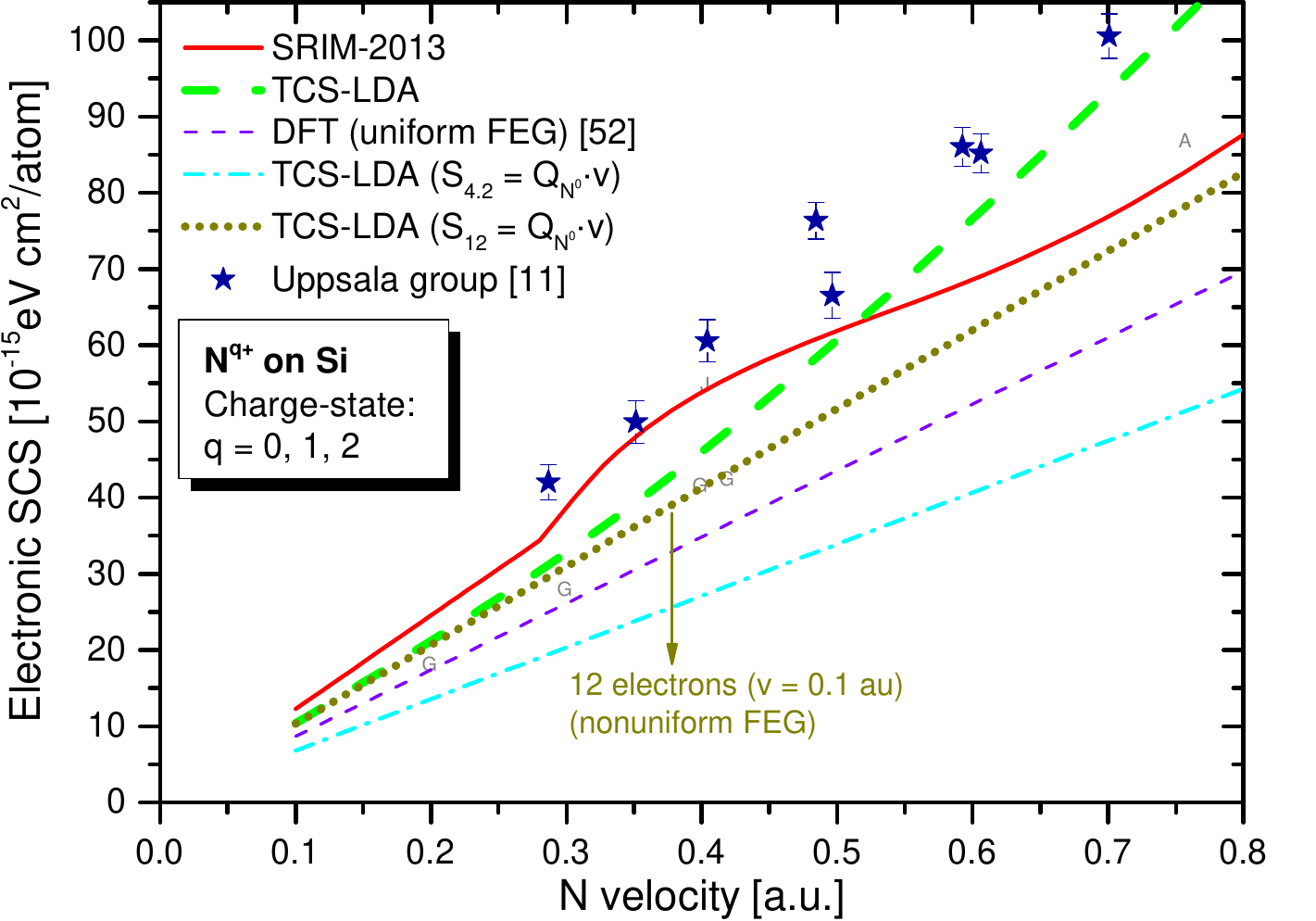}
\caption{As in Fig.~\ref{fig:He_Si_56keV_LDA}, we presented the electronic SCS averaged over the charge states for N$^{\text{q}+}$ ions on Si as a velocity function. TCS-LDA results for the linear SCS: the dark yellow short-dot line is the result with full ELF, in which 12 electrons effectively contribute to the nonuniform FEG in electron-N$^0$ interaction.}
\label{fig:N_Si_16keV_LDA}
\end{center}
\end{figure}

\begin{figure}[h]
\begin{center}
\includegraphics[angle=0,width=9cm,clip]{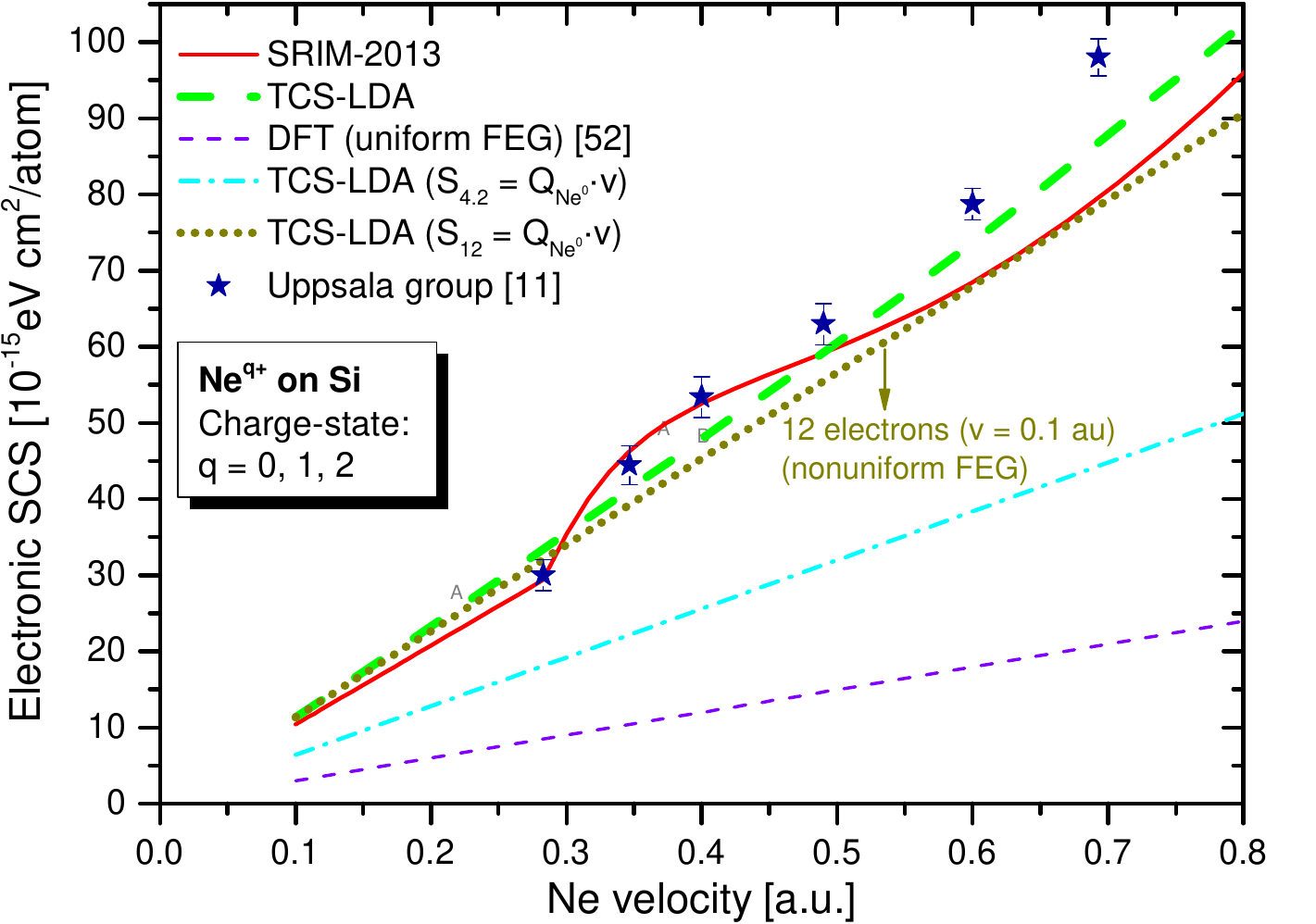}
\caption{As in Fig.~\ref{fig:He_Si_56keV_LDA}, electronic SCS averaged over the charge states for Ne$^{\text{q}+}$ ions on Si as a velocity function is plotted. TCS-LDA results for the linear SCS: the dark yellow short-dot line is the result with full ELF, in which 12 electrons effectively contribute to the nonuniform FEG in electron-Ne$^0$ interaction.}
\label{fig:Ne_Si_16keV_LDA}
\end{center}
\end{figure}

The electronic SCS of H, He, N, and Ne in Si are presented in Figs.~\ref{fig:H_Si_10000keV_Penn} to \ref{fig:Ne_Si_16keV_LDA}. The letters represent the experimental IAEA database.~\cite{iaea} Furthermore, we have included recent data from the Uppsala group, \cite{ntemou2023} represented by star symbols. In Fig.~\ref{fig:H_Si_10000keV_Penn}, the black open circle symbols represent the experimental data from Ref.~\cite{arista1999} The red solid line represents the semi-empirical predictions of SRIM-2013. \cite{srim2010} We employed three non-perturbative methods 
%for the theoretical analysis of 
to compute ion's SCS: real-time TDDFT, TCS, and IDA, %where X $\equiv$ 
combined with Penn or LDA. The results obtained using Eqs.~\eqref{eq:lda} and \eqref{eq:penn} are summarized in Fig.~\ref{fig:H_Si_10000keV_Penn}. The results show that for $v>1.0$ a.u. (25 keV), the real-time TDDFT-Penn and IDA-Penn approaches are highly similar and both are in excellent agreement with experimental data, especially with those from the Uppsala group. \cite{ntemou2023} Nevertheless, for velocities between $2.0$ and $3.2$ a.u., the real-time TDDFT-Penn approach agrees better with the Uppsala data.

%In Fig.~\ref{fig:H_Si_10000keV_Penn}, the results of the SCS calculations using the Penn and LDA approaches are presented. The results show that for $v>1.0$ a.u. (25 keV), the real-time TDDFT-Penn and IDA-Penn approaches are highly similar, and both are in excellent agreement with experimental data, especially those from the Uppsala group \cite{ntemou2023}. Nevertheless, for velocities between $2.0$ and $3.2$ a.u., the real-time TDDFT-Penn approach agrees better with the Uppsala data.

For $v<1.0$ a.u., the real-time TDDFT-Penn approach tends to overestimate the stopping power. One plausible explanation for this discrepancy is to consider the inclusion of the neutral H$^0$ charge state. In this velocity range, this charge state is predominant. For $v < 0.6$ a.u., the CasP 6.0 estimation indicates a 100\% probability of H$^0$ presence. \cite{casp2021,matias2017} Therefore, these observations suggest the importance of considering the appropriate charge state for this velocity range in electronic stopping power calculations.

As observed in Fig.~\ref{fig:H_Si_10000keV_Penn}, it is evident that the TCS-Penn results are lower than the actual values in the Bragg peak region. At high velocities, the TCS-Penn results slowly converge to the experimental results. This is a fundamental feature of the TCS method, as it does not precisely converge to the Bethe formula, which has been discussed in previous studies. \cite{grande16,matias2017}

The results obtained from real-time TDDFT-LDA, TCS-LDA, and IDA-LDA show that the LDA tends to overestimate the SCS compared to the average values of the experimental data
%The application of LDA tends to overestimate the stopping power in the approximately 
in the velocity range $v_{\rm F} < v < 5$ a.u., where $v_{\rm F} = 1.919/r_s$ a.u.

The Penn approach combined with the real-time TDDFT and IDA methods provides accurate results in agreement with the experimental data. The IDA-Penn results show excellent agreement at all velocities in Fig.~\ref{fig:H_Si_10000keV_Penn}. For proton velocities lower than the Fermi velocity, 
%($v < v_{\rm F}$ a.u.), 
the agreement of TCS-LDA with the experimental data is similar to that of IDA-Penn. Despite this similarity in the results, it is expected that TCS-LDA is more accurate than IDA-Penn. \cite{grande16,matias2017} Although IDA has shown excellent agreement with the experimental data, its use in electronic stopping power estimates for $Z>2$ leads to underestimated results. \cite{matias2017} For this reason, only TCS results are present for $Z > 1$ and $v \le 0.8$ a.u.

Real-time TDDFT results obtained from plane-wave pseudopotential calculations and scaled by a velocity-dependent factor to account for the absent core electrons \cite{kanai:2017} agree with our IDA-LDA and TDDFT-LDA results at $v>2$ a.u. However, the results of Yost \emph{et al.} \cite{kanai:2017} show the maximum of the stopping shifted to higher velocity as compared to our results and closer to IDA-Penn and TCS-LDA at velocities below 2 a.u.

In Fig.~\ref{fig:H_Si_625keV_LDA}, we display the same as in Fig.~\ref{fig:H_Si_10000keV_Penn} but in linear scale and including the original nonlinear DFT estimates to SCS,~\cite{ECHENIQUE1990229} which are 
proportional to velocity $dE/dz\equiv S=Q\cdot v$, where $Q$ is a velocity-independent friction coefficient by Echenique \textit{et al.}~\cite{echenique1986} The DFT-LDA calculations for the friction coefficients are expected to agree with our calculations (TCS-LDA) at $v\to 0$. Still, the velocity-proportional stopping extrapolation can overestimate or underestimate the stopping power for finite ion velocities. This will be important when we analyze the SCS for heavier ions at low velocities.

The predominant charge states for He, N, and Ne projectiles at $v<v_F$ are $q=0, 1$ for He and $q=0, 1, 2$, and $3$ for N and Ne. Although the real-time TDDFT method is considered a benchmark, it was not used in calculating the electronic stopping of He, N, and Ne in Si because the current code considers only bare ions. 

In Figs.~\ref{fig:H_Si_625keV_LDA}-\ref{fig:Ne_Si_16keV_LDA}, the linear electronic SCS was determined from the friction coefficient values $Q$. The stopping power values were obtained using the TCS-LDA approach (Eq.~\ref{eq:lda}) in $v=0.1$ a.u. CasP 6.0 software estimates show that the projectiles have neutral charges at this speed. Therefore, we consider only the charge state $q=Z-n_b=0$, where $n_b$ represents the number of electrons bound to the ion. The values of $Q$ for H$^0$, He$^0$, N$^0$, and Ne$^0$ are $0.2167$, $0.4035$, $1.0022$, and $1.0458$ a.u., respectively.

The DFT results presented in Figs.~\ref{fig:H_Si_625keV_LDA}-\ref{fig:Ne_Si_16keV_LDA} were obtained assuming a uniform FEG for the Si valence band. \cite{ECHENIQUE1990229} We used a WS radius of $r_s = 1.97$ a.u. for this uniform FEG obtained from the experimental plasmon energy of Si.~\cite{isaacson:1975,brandt:1981} Density parameter $r_s=1.97$ a.u. was used in this DFT calculation, corresponding to $4.2$ valence electrons. DFT results (violet dashed line) show that the assumption of uniformity in electron density fails for electronic SCS estimates of N and Ne in Si as pointed out in the experimental work. \cite{ntemou2023}

When considering electron nonuniformity in TCS-LDA calculations using  $1.03 \leq r \leq 3.195$ range, as depicted in Fig.~\ref{fig:electrons-number}, a notable agreement with experimental data is observed in the electronic SCS estimation, as shown by the cyan dash-dotted line in Figs. \ref{fig:H_Si_625keV_LDA}-\ref{fig:Ne_Si_16keV_LDA}. 

Based on the electron density ranges, as shown in Fig.~\ref{fig:electrons-number}, we can draw the following conclusions: TCS-LDA calculations suggest that H$^0$ projectiles interact with a nonuniform FEG of 6 electrons, which is higher than the 4.2 electrons prediction (uniform FEG).~\cite{brandt:1981} Similarly, He$^0$ projectiles interact with 10 electrons, while N$^0$ and Ne$^0$ projectiles interact with 12 electrons from Si. As the projectile nuclear charge increases, this discrepancy may stem from increased electron-hole pair excitations from inner shells.~\cite{ntemou2023} However, this phenomenon is not adequately described in FEG-based models due to the absence of binding energy in the model, which rapidly diminishes as the velocity approaches zero. Another noteworthy phenomenon is the electron elevator mechanism observed in \textit{ab initio} calculations for Si in Si, as discussed in Ref.,\cite{Lim2016} where excitations across energy gaps via a dynamical gap state are present.

The dark yellow short-dotted lines in Figs.~\ref{fig:H_Si_625keV_LDA}-\ref{fig:Ne_Si_16keV_LDA} represent the TCS-LDA results for the effective electrons in a nonuniform FEG. Therefore, it is crucial to consider electron nonuniformity in these calculations, as it significantly improves electronic stopping and allows for the emulation of effects beyond the simple electron-hole excitation.

SCS results for He, N, and Ne are presented in Figs.~\ref{fig:He_Si_56keV_LDA}, \ref{fig:N_Si_16keV_LDA}, and \ref{fig:Ne_Si_16keV_LDA}, respectively. The TCS-LDA result (green-dashed line) for He shows a good agreement with the experimental data.~\cite{iaea,ntemou2023} The slope coefficient of the linear SCS (dark yellow short-dot line) provided by TCS-LDA is much more accurate than the uniform nonlinear DFT results. 

\noindent
Note that this discrepancy does not come from nonlinear DFT but rather from the assumption of uniformity in the electron density. The results for Ne in Si have the greatest discrepancy between the results with uniform and nonuniform electron density, as shown in Fig. \ref{fig:Ne_Si_16keV_LDA}.

\begin{figure}[h]
\begin{center}
\includegraphics[angle=0,width=9cm,clip]{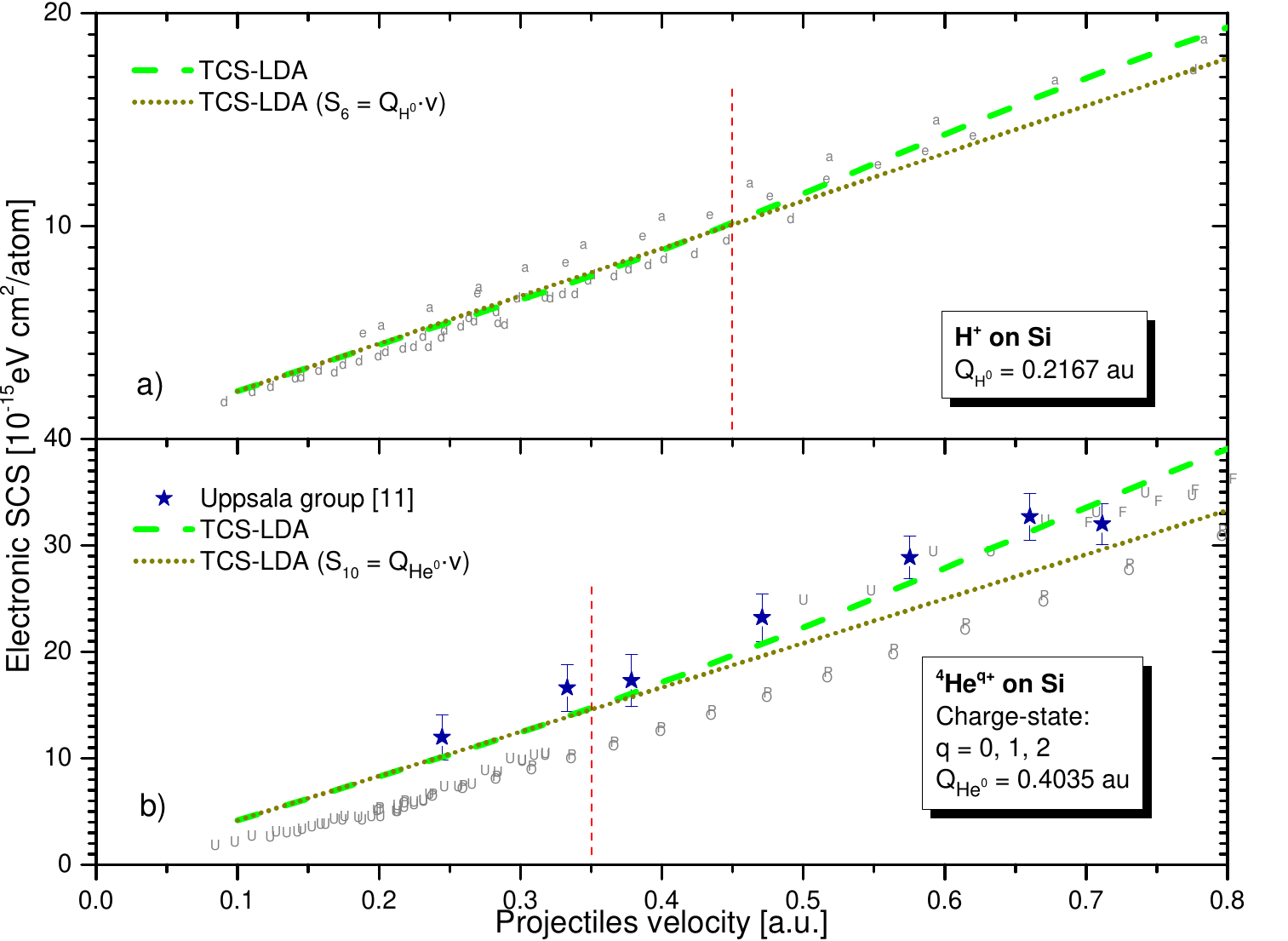}
\caption{Electronic SCS results for the H$^0$ (a) and He$^0$ (b). The green-dash line is the result calculated in the TCS-LDA approach (nonlinear SCS to velocity); the dark yellow short-dot line is the result of the linear SCS ($S = Q\cdot v$) performed in the TCS-LDA approach at $v = 0.1$ a.u. The uppercase letters are experimental data available at the IAEA database~\cite{iaea}, and the royal star symbols are Uppsala group results.~\cite{ntemou2023}}
\label{fig:stp_linear_nonlinear_H-He}
\end{center}
\end{figure}

As evident from Figs.~\ref{fig:H_Si_625keV_LDA}-\ref{fig:Ne_Si_16keV_LDA}, the dependence of electronic stopping on the nonuniformity of the electron density increases as the atomic charge of the ion increases (at $v = 0.1$ a.u.). The consequences of these effects will be discussed in relation to Figs. \ref{fig:stp_linear_nonlinear_H-He} and \ref{fig:stp_linear_nonlinear_H-He}.

%The latter is calculated from the stopping power coefficient, as explained earlier. 

\begin{figure}[h]
\begin{center}
\includegraphics[angle=0,width=9cm,clip]{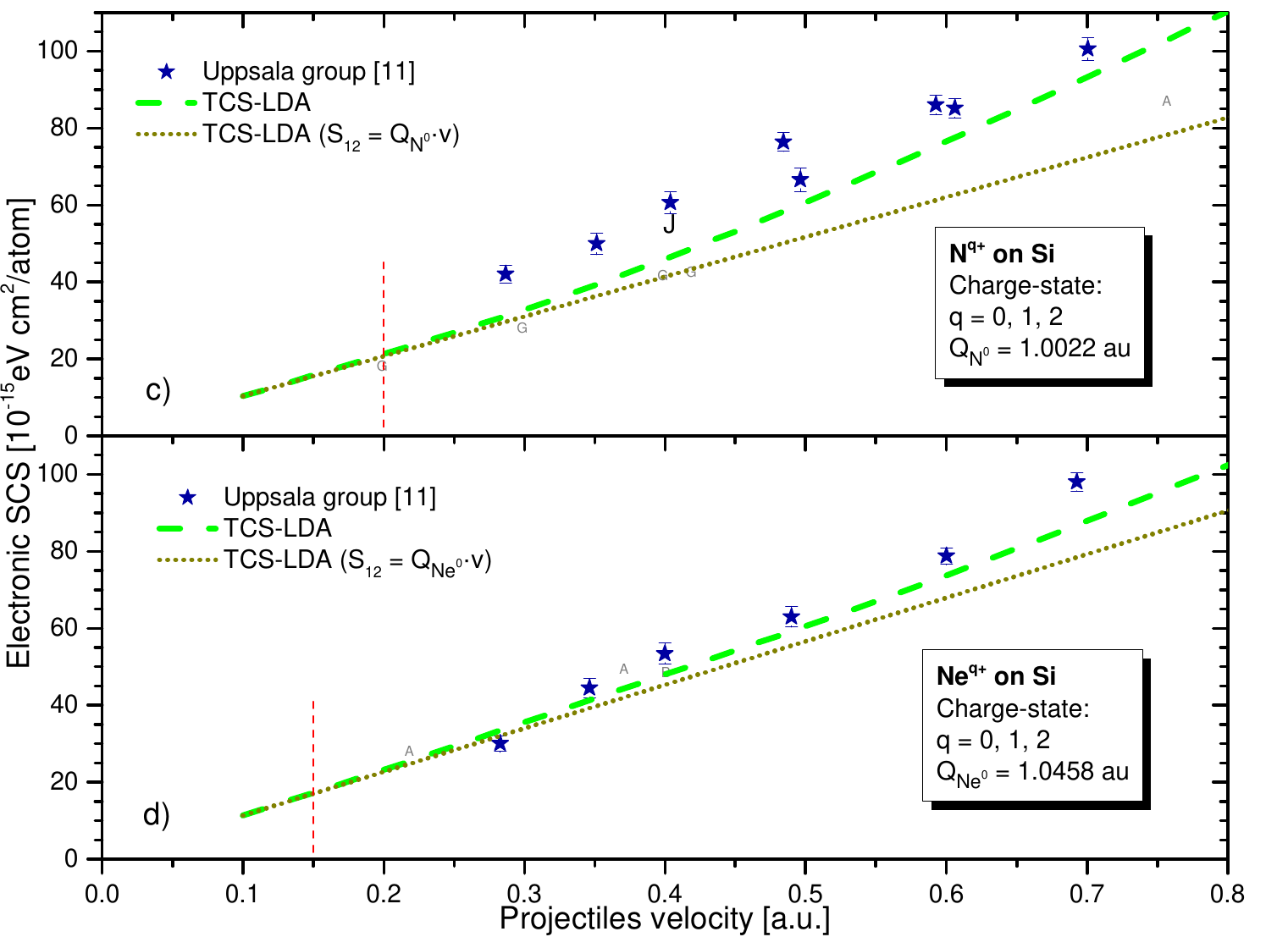}
\caption{As in Fig.~\ref{fig:stp_linear_nonlinear_H-He}, electronic SCS results for the N$^0$ (c) and Ne$^0$ (d) are presented.}
\label{fig:stp_linear_nonlinear_N-Ne}
\end{center}
\end{figure}

Figs.~\ref{fig:stp_linear_nonlinear_H-He} and \ref{fig:stp_linear_nonlinear_N-Ne} show the comparison between the SCS (nonlinear to velocity) calculated with the TCS-LDA (Eq.~(\ref{eq:lda})) and the SCS (linear to velocity) calculated from the stopping coefficient ($S = Q\cdot v$), also performed in the TCS-LDA approach; we calculate $Q$ considering the value of the stopping power at $v = 0.1$, then we extrapolate velocity-proportional electronic stopping for $v>0.1$ a.u.

%\noindent
We note a velocity-proportionality breakdown in SCS that occurs at velocities of $0.45$ [for H, Fig.~\ref{fig:stp_linear_nonlinear_H-He}(a)], $0.35$ [for He, Fig.~\ref{fig:stp_linear_nonlinear_H-He}(b)], $0.20$ [for N, Fig.~\ref{fig:stp_linear_nonlinear_N-Ne}(c)], and $0.15$ a.u. [for Ne, Fig.~\ref{fig:stp_linear_nonlinear_N-Ne}(c)]. Moreover, we observe a pattern for these breaks: the linear SCS range decreases as the atomic charge of the ion increases. This indicates that the velocity range is influenced solely by weakly bound electrons from the valence band, which contributes to the stopping and becomes narrower with increasing projectile charge. The uniform FEG approach fails to consider all potential excitations. Deeper electrons can be dynamically excited through alternative mechanisms, such as ``elevator'' and/or ``promotion,'' enabling physical scenarios where a localized free electron gas can emulate ion energy loss. This aligns with the observations from Ref.\cite{Lim2016,ntemou2023}.

\begin{figure}[h]
\begin{center}
\includegraphics[angle=0,width=9cm,clip]{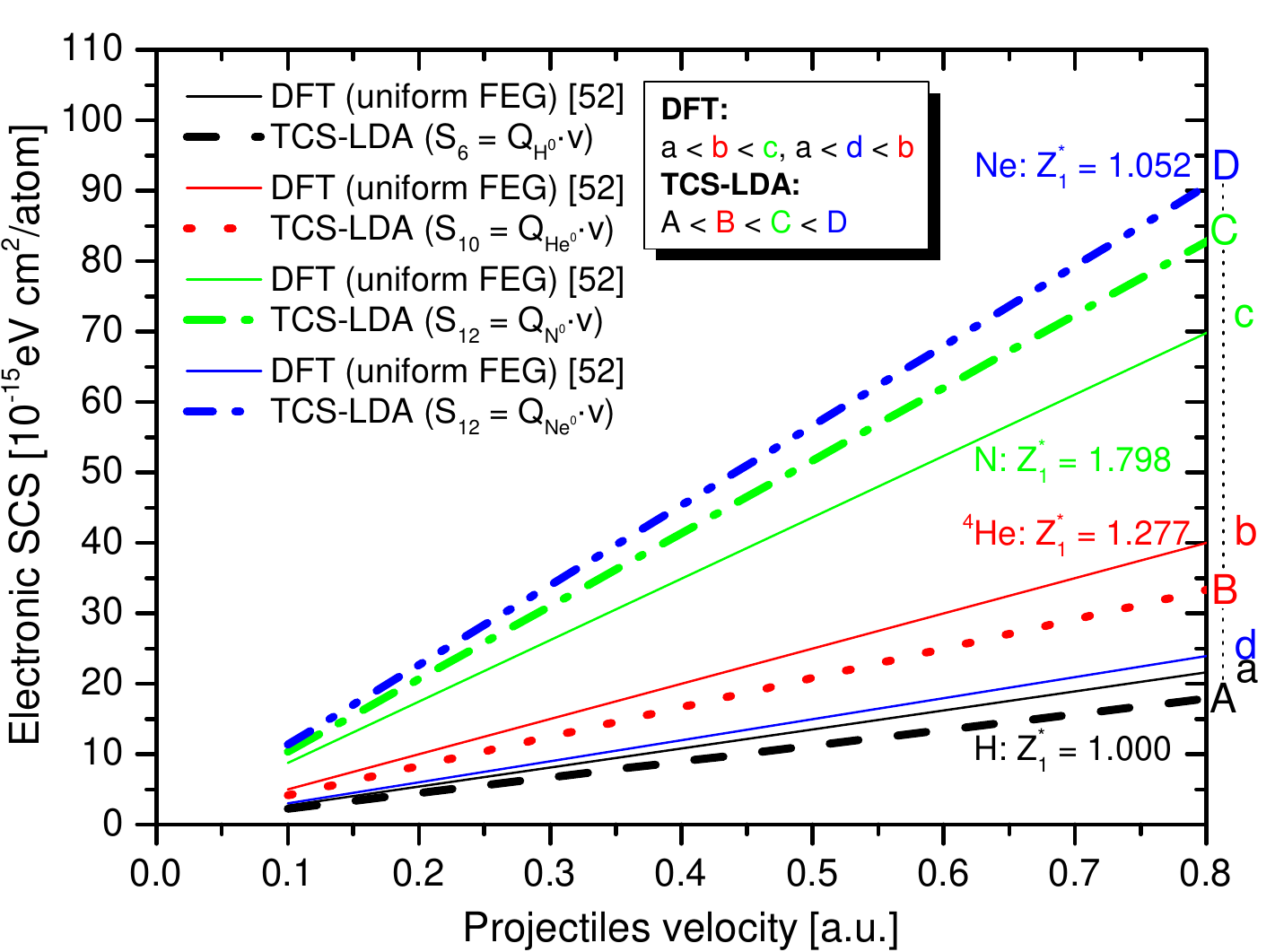}
\caption{Electronic SCS results for the H$^0$, He$^0$, N$^0$, and Ne$^0$. The non-continue lines are the results of the electronic stopping coefficient ($S = Q\cdot v$) performed in the TCS-LDA approach at $v = 0.1$ a.u.: H$^0$ (black dash line, letters A), He$^0$ (red dot line, letters \textcolor{red}{B}), N$^0$ (green dash-dot line, letters \textcolor{green}{C}), and Ne$^0$ (blue dash-dot dot line, letters \textcolor{blue}{D}). DFT results~\cite{ECHENIQUE1990229} are shown for comparative proposal: H$^0$ (black solid line, letters a), He$^0$ (red solid line, letters \textcolor{red}{b}), N$^0$ (green solid line, letters \textcolor{green}{c}), and Ne$^0$ (blue solid line, letters \textcolor{blue}{d}). $Z_1^*$ represents the effective charge values for each neutral atom.~\cite{ECHENIQUE1990229}}
\label{stp_linear-nonlinear_comp_H-He-N-Ne}
\end{center}
\end{figure}

In Fig.~\ref{stp_linear-nonlinear_comp_H-He-N-Ne}, an interesting result is presented, which is similar to Figs.~\ref{fig:H_Si_625keV_LDA}-\ref{fig:Ne_Si_16keV_LDA} in terms of DFT and TCS-LDA estimates.
The uniform DFT results (solid lines) are presented for H$^0$ (a), He$^0$ (\textcolor{red}{b}), N$^0$ (\textcolor{green}{c}), and Ne$^0$ (\textcolor{blue}{d}). On the other hand, TCS-LDA results for the linear SCS are shown as A (H$^0$), \textcolor{red}{B} (He$^0$), \textcolor{green}{C} (N$^0$), and \textcolor{blue}{D} (Ne$^0$).
%
%\noindent
In uniform nonlinear DFT predictions, we observed a relationship between the inclinations of the SCS lines. Specifically, we have found that a $<$ \textcolor{red}{b} $<$ \textcolor{green}{c} and a $<$ \textcolor{blue}{d} $<$ \textcolor{red}{b}. The figure displaying the SCS results also includes the effective charge values for each projectile.~\cite{ECHENIQUE1990229} It is worth noting that the effective charge values increase from H$^0$ to N$^0$, but the effective charge value of Ne$^0$ is slightly higher than that of H$^0$. These oscillations in effective charge values of atoms are well-known and can be confirmed in Fig.~15 of Ref.~\cite{ECHENIQUE1990229}.

When we consider the nonuniform FEG and account for the inner shells' electron contributions, we notice that the slopes of the SCS lines increase as the ions' atomic charge increases. This nonuniformity electron density and the contributions of inner electrons to SCS alter the slope pattern: A $<$ \textcolor{red}{B} $<$ \textcolor{green}{C} $<$ \textcolor{blue}{D}.

\section{Conclusions}

%The TDDFT-Penn results demonstrated excellent agreement with the experimental data at velocities above approximately 1.0 au; however, there was a tendency to overestimate the SCS below this threshold. This was attributed to the necessity of considering including the neutral H$^0$ charge state.\textcolor{blue}{Melhor cortar essa parte e focar a conclusão nas novidades!}

%On the other hand, IDA-Penn results proved highly accurate for protons in Si, exhibiting excellent agreement across all velocities. On the contrary, the TCS-Penn approach underestimated the experimental data in the Bragg peak region, gradually converging at higher velocities. Meanwhile, the TDDFT-LDA and IDA-LDA approaches tended to overestimate the SCS, especially at velocities between $v_F < v < 6$ a.u., approximately.\textcolor{blue}{Melhor cortar essa parte e focar a conclusão nas novidades!}

Our findings emphasize the importance of considering nonuniform electron density, shedding new light on local density approximations, particularly their ability to dynamically incorporate contributions from deeper-band electrons. Additionally, we observe a breakdown in the proportionality relationship between stopping power and velocity as $v$ approaches zero. The larger the projectile charge, the lower the velocity at which this breakdown occurs. This phenomenon, explaining the findings observed in Ref., \cite{ntemou2023} likely arises from the possibility of dynamic changes in deep energy levels due to ``elevator'' and/or ``promotion'' mechanisms. These mechanisms can create scenarios resembling a local free electron gas.

While the standard DFT approach proved inadequate for these calculations, a nonuniform DFT approach, or ideally a full \textit{ab initio} calculation, is expected to yield results consistent with the TCS-LDA estimates for electronic stopping.
Using local density approximation contributes to a better understanding of the complexities in electron-ion interactions in this context.  

\begin{acknowledgments}
This work was partially supported by IPEN (project number 2020.06.IPEN.32) and CNPq (project number 406982/2021-0). The authors acknowledge FAPESP for supporting the computer cluster (process numbers 2012/04583-8 and 2020/04867-2). TFS acknowledges the financial support provided by CNPq-INCT-FNA (project number 464898/2014-5).
\end{acknowledgments}

\section*{AUTHOR DECLARATIONS}

\subsection*{Conflict of Interest}
The authors have no conflicts to disclose.

\subsection*{Author Contributions}

F. Matias: Conceptualization, Methodology, Software, Validation, Writing - Original Draft. P. L. Grande: Methodology, Validation, Writing - Review \& Editing. N. E. Koval: Data Curation, Writing - Review \& Editing. J. M. B. Shorto: Funding acquisition, Data Curation, Writing - Review \& Editing. T. F. Silva: Funding acquisition, Methodology, Validation, Writing - Review \& Editing. N. R. Arista: Methodology, Validation, Writing - Review \& Editing. All authors reviewed the paper.

\section*{Data Availability}

The data that support the findings of this study are available from the
corresponding author upon reasonable request.

\nocite{*}
\bibliography{biblio}% Produces the bibliography via BibTeX.

\end{document}